\documentclass[aps,prl,twocolumn,showpacs,preprintnumbers]{revtex4}
\usepackage{graphics}
\def \beq{\begin{equation}}
\def \eeq{\end{equation}}
\def \beqa{\begin{eqnarray}}
\def \eeqa{\end{eqnarray}}
\def \md{m_{\scriptscriptstyle D}}
\def \mt{m_{\scriptscriptstyle T}}
\def \lms{\Lambda_{\overline{\scriptscriptstyle MS}}}

\begin{document}

\title{Does the QCD plasma contain propagating gluons?}
\author{Saumen \surname{Datta}}
\email{saumen@Physik.Uni-Bielefeld.de}
\affiliation{Fakult\"at f\"ur Physik, Universit\"at Bielefeld,
         D-33615 Bielefeld, Germany.}
\author{Sourendu \surname{Gupta}}
\email{sgupta@tifr.res.in}
\affiliation{Department of Theoretical Physics, Tata Institute of Fundamental
         Research,\\ Homi Bhabha Road, Mumbai 400005, India.}

\begin{abstract}
Comparison of two appropriately chosen screening masses of colour singlet
operators in the pure glue QCD plasma indicates that at sufficiently
high temperature it contains a weakly-interacting massive quasi-particle
with the quantum numbers of the electric gluon. Still in the deconfined
phase, but closer to $T_c$, the same mass ratio is similar to that at zero
temperature, indicating that the propagating modes are more glueball-like,
albeit with a lower scale for the masses. We observe a continuity between
these two regimes.
\end{abstract}
\pacs{11.15.Ha, 12.38.Gc \hfill TIFR/TH/02-25, BI-TP 2002/17}
\preprint{TIFR/TH/02-25, BI-TP 2002/17, hep-lat/0208001}
\maketitle

With the RHIC fully operational and busy taking data, certain questions
about the treatment of the QCD plasma have become urgent. One of the most
basic is about the modes of excitation in a plasma: does it contain weakly
interacting quark and gluon quasi-particles, or some more complicated
collective excitations?  We show here that lattice computations yield
detailed answers to this question.

The Debye screening mass, $\md$, in pure gauge QCD, at temperature $T>T_c$
has an expansion in the strong coupling $g$ of the form---
\beq
   \frac{\md}T = g - \frac3{4\pi}g^2\log g + bg^2 + cg^3 + {\cal O}(g^4)
\label{pert}\eeq
where the leading term $g$ is well known, the second term has been
extracted in perturbation theory \cite{toni} and the non-perturbative
coefficients $b = 2.46\pm0.15$ and $c = -0.49\pm0.15$ have been computed
in a lattice simulation of dimensionally reduced QCD \cite{kari}. The
QCD coupling $g$ has to be evaluated at the scale $6.742T$. Since
$T_c/\lms=1.15\pm0.05$ \cite{scale}, $g={\cal O}(1)$ for $T/T_c\simeq3$
or less, and the error due to the neglect of the $g^4$ term is about
$35d$\%, where $d$ is the coefficient of this term. As a result, it
becomes difficult to validate perturbation theory by comparing this
screening mass to lattice data \cite{olaf}.

In this work we address a prior question: at temperatures of interest to
current and near-future experiments what mediates the longest correlations
in the plasma?  We examine this by comparing screening masses, $m$,
obtained from correlations of two gauge invariant operators with different
symmetry properties.  They are chosen in such a way that one would be
obtained by the exchange of two electric gluons while the other would need
three, if indeed such gluons are the lightest excitations in the plasma.
As a result, the two screening masses would be roughly in the ratio 3/2.

Since screening masses involve the transfer matrix in a spatial direction,
they are classified by the symmetry group of the lattice sliced
perpendicular to a spatial direction \cite{bernd,saumen,arnold}. For
the thermodynamics of a 3+1 dimensional field theory realised on a
hypercubic Euclidean lattice, since the Euclidean time direction is
distinguished from the spatial directions, this is the group $D_4\times
Z_2(T)\times Z_2(C)$, where $D_4$ is the tetragonal group, $Z_2(C)$
is the charge-conjugation symmetry of the fields, and the $Z_2(T)$
factor arises from the symmetry $t\leftrightarrow-t$ \cite{bernd,saumen}.
The transfer matrix can be block diagonalised in irreps of this group.

Extensive thermodynamic quantities depend only on the lowest eigenvalue
of the transfer matrix, and the phase structure is determined by the
degeneracies and symmetries of the corresponding eigenvectors. In high
temperature QCD, these are the scalar, $A_1^{++}$ irrep \cite{irreps}.
In the confined phase, the unique ground state is scalar under the $Z_3$
symmetry of the center of the colour $SU(3)$ group.  In the deconfined
phase the ground state is three-fold degenerate and corresponds to
the three irreps of $Z_3$. On any finite lattice the degeneracy is
lifted by an exponentially small quantity due to tunnelings between
these states. In the $A_1^{++}$ sector, therefore, an unphysical small
``tunneling'' mass, $\mt$, may dominate the screening. We show later
that we can control this and obtain the correct physical screening mass.

Since the link variable, $U\sim\exp[\int dx igA]\sim\exp[iagA]$ (here $a$
is the lattice spacing), one can turn this around and define a lattice
gluon field of momentum $k$ by the relation
\beq
   A_\mu(k) = i\sum_x{\rm e}^{ik\cdot x}\left[
    U_\mu(x)-U^\dag_\mu(x)-{\rm Im}\,{\rm Tr}\,U_\mu(x)\right],
\label{gluon}\eeq
where the sum is over all lattice sites in a slice, the components of $k$
run over the set $2\pi l/N$ where $1\le l\le N$ and $N$ is $N_t$
for the temporal momenta and $N_s$ for the spatial momenta (we assume
an $N_t\times N_s^3$ lattice).  This definition gives an element of
the $SU(N)$ algebra which goes over into the continuum definition of
the colour octet gluon field in the limit of zero lattice spacing.
The electric gluon $A_{\hat t}(0)$ is in the $A_2^{--}$ irrep, Since
$A_2^{--}\otimes A_2^{--} = A_1^{++}$ and $A_2^{--}\otimes A_2^{--}\otimes
A_2^{--} = A_2^{--}$, screening correlators in the colour singlet
$A_1^{++}$ and $A_2^{--}$ sectors would be dominated by two and three
electric gluon exchange respectively. We discuss the influence of the
magnetic sector later.

In this work we use two classes of $A_1^{++}$ operators--- Wilson loops,
specifically several linear combinations of the plaquette and the planar
6-link loop (sometimes called the fenster), and the trace of the real
part of the Wilson line. For the $A_2^{--}$ operator we use the imaginary
part of the trace of the Wilson line.  The zero momentum projection is
obtained as usual by summing over all sites in the slice.  Since the
Wilson lines are non-trivial under $Z_3$ transformations, they are
evaluated in the phase where the expectation value is real \cite{rotate}.

Due to the $Z_3$ symmetry of the vacua, there are only two tunneling
masses of relevance--- one in the vicinity of $T_c$ due to tunneling
between the disordered state and any of the ordered states, and the other
for all $T\ge T_c$ due to tunneling between any two of the ordered states
(the latter are relevant only to operators which are non-trivial under
$Z_3$). Each mass has very specific dependence on the volume, $\overline
V=a^3N_s^2N_t$, of a slice orthogonal to the direction of propagation---
\beq
   \mt(\overline V) = \left(\frac C{\overline V^\alpha}\right)
        \exp(-\sigma\overline V),
\label{tunnel}\eeq
where $C$, $\alpha$ and $\sigma$ are constants. In a d-dimensional scalar
theory a one-loop computation gives $\alpha=d/2$ \cite{zinn}. Tunneling
arises when, on a finite system, simulations start exploring the
non-Gaussian part of the free energy away from local minima. Clearly,
an appropriate correlation function can give $\mt$ only when the order
parameter distribution shows multiple peaks.  A consequence of eq.\
(\ref{tunnel}) is that one can perform a finite size scaling study to
check whether the lowest screening mass obtained is a tunneling mass.

\begin{table}[htbp]
  \begin{center}\begin{tabular}{|c|c|l|c|}  \hline
  $N_s$ & $\beta$ & $T/T_c$ & statistics \\
  \hline
  24 & 5.6500 & 0.89 (1) & 11370 \\
  24 & 5.6800 & 0.97 (1) & 12080 \\
% 24 & 5.6850 &          & 10000$^*$ \\ 
  16 & 5.6908 & 1.00     & 20000 \\
  20 & 5.6918 & 1.00     & 20000 \\
  24 & 5.6920 & 1.00     & 20000 \\
  16 & 5.7010 & 1.02 (1) & 10000 \\
  20 & 5.7010 & 1.02 (1) & 10000 \\
  24 & 5.7010 & 1.02 (1) & 20020 \\
  24 & 5.7100 & 1.04 (1) & 10160 \\
  24 & 5.7200 & 1.07 (1) & 18010 \\
% 24 & 5.7300 &          & 10000$^*$ \\
% 24 & 5.7500 &          & 8000 \\
  24 & 5.8000 & 1.27 (2) & 32220 \\
  24 & 5.8941 & 1.50     & 15130 \\
  24 & 6.0625 & 2.00     & 31350 \\
  24 & 6.3500 & 3.00     & 31630 \\
  \hline
  \end{tabular}\end{center}
  \caption[dummy]{For these runs on $N_t=4$ lattices, measurements were
     taken every 5th sweep, where every sweep was done with a 3 hit pseudo
     heat-bath (except for the runs at $T_c$, where measurements were taken
     every 10th sweep, each with a 6 hit pseudo heat-bath). In each case,
     an initial 1000--3000 sweeps were discarded for thermalization, and
     masses were extracted by a jack-knife analysis with 100 bins.}
\label{tb.runs}\end{table}

Details of our runs with the pure gauge Wilson $SU(3)$ action are
summarized in Table \ref{tb.runs}.  The critical coupling, $\beta_c$,
and its shift on finite lattices, is known for $N_t=4$ with high
precision \cite{crit,entropy}.  Since the finite-size shift of $\beta_c$
on the smallest lattice, $N_s=16$, is less than 2 parts in $10^3$, the
temperature scale is known with high precision at $T_c$.  It is also known
at similar precision at $1.5T_c$, $2T_c$ and $3T_c$ from measurements with
$N_t=6$, 8 and 12. At other points the temperature scale is interpolated
through the QCD beta function and has errors \cite{scale}, which are
indicated in the table.  Loop operators are measured at five levels of
single link fuzzing \cite{saumen}.  Cross correlations between all loops
in the same irrep are measured and the lowest screening mass obtained
by a variational procedure. Other details of measurements and analysis
remain as in \cite{saumen}.

\begin{figure}[htb]\begin{center}
   \scalebox{0.6}{\includegraphics{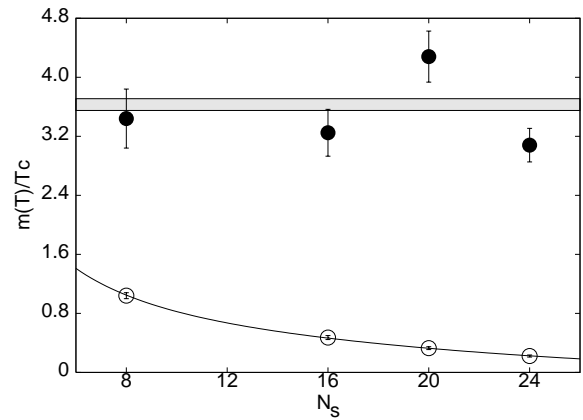}}
   \end{center}
   \caption{$A_1^{++}$ screening masses at $T_c$ and their dependence on the lattice
      size. The lowest mass (unfilled circles) is a tunneling mass, as evidenced by
      the good fit to the form in eq.\ (\ref{tunnel}). Also shown are estimates of
      the second variational mass (filled circles) and a comparison with a measurement
      of the lowest mass at $T/T_c=0.97$ on a $4\times24^3$ lattice (horizontal band).
      Data for $N_s=8$ is from \cite{saumen}.}
\label{fg.trend}\end{figure}

At $T=T_c$, the statistics collected are large enough that tunnelings
between the deconfined and confined phases occur many times, as do
those between different deconfined phases.  As a result, it is only to
be expected that the lowest screening mass that we can extract in the
$A_1^{++}$ sector is the tunneling mass. Evidence for this is the good fit
to eq.\ (\ref{tunnel}) shown in Figure \ref{fg.trend}; the fit gave
$\chi^2=1.005$ per degree of freedom.

To stabilise the pure phases we move away from $T_c$. The distance
from criticality, $\Delta\beta=|\beta-\beta_c|$, needed to remain in a
single phase for arbitrarily long runs depends on $N_s$: $\Delta\beta$
can decrease exponentially with $N_s$.  In the confined phase we
have estimated $m(T_c)$ by a measurement with $\Delta\beta=0.0125$,
corresponding to a $3\pm1$\% shift of $T$ below $T_c$. A $4\times24^3$
lattice was observed to stay in the confined phase throughout the run. The
second variational level was seen to correspond to the screening
mass in this phase. Our measurement of the $A_1^{++}$ screening
mass at $T/T_c=0.97$ gives $m(T)/T_c=3.41\pm0.08$ (at $T/T_c=0.89$
we have $m(T)/T_c=3.52\pm0.07$).  In comparison, a zero temperature
measurement of the scalar glueball mass at a similar lattice spacing
gives $m(T=0)/T_c=3.93\pm0.05$ \cite{tzero}.  At $T=0$, glueball masses
depend strongly on the lattice spacing, whereas mass ratios are less
sensitive. In this phase, we may then expect that the ratio $m(T)/m(T=0)$
may scale better than the ratio $m(T)/T$ as the lattice spacing goes
to zero. Our measurements show that $m(T)/m(T=0)\sim0.8$ in the scalar
channel near $T_c$, in agreement with recent measurements from temporal
correlators \cite{ishii}.

Another method of extracting the physical correlation length is to pick
out the configurations in which the whole lattice is in a single phase. We
do this through the distribution of the action density, which has peaks
corresponding to each of the phases.  By restricting measurements
of correlation functions to configurations with some $S>S_{cut}$,
suitably chosen, we can isolate the deconfined phase and measure the
physical screening mass.  This procedure on the largest lattice gave
$m(T_c)/T_c\approx1$ in the deconfined phase.  However with increasing
$S_{cut}$ statistics become poorer; consequently, errors increase rapidly,
and it is hard to quote a more precise value.

\begin{figure}[hbt]\begin{center}
   \scalebox{0.6}{\includegraphics{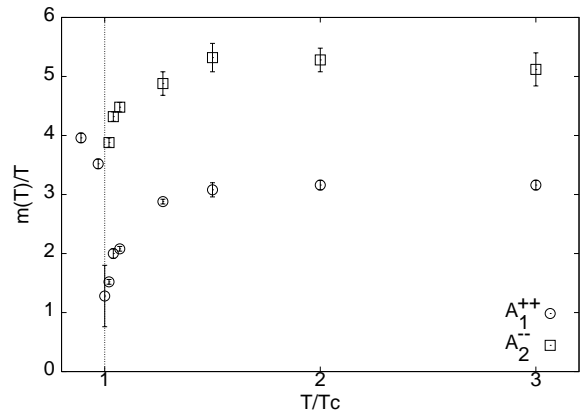}}
   \end{center}
   \caption{$A_1^{++}$ and $A_2^{--}$  screening masses as a function of
      temperature on $4\times24^3$ lattices. Note the discontinuity in
      the $A_1^{++}$ screening mass at $T_c$.}
\label{fg.a1p}\end{figure}

We worked in the deconfined phase by taking $\Delta\beta\ge0.0085$ for
$N_s=16$, 20 and 24, corresponding to moving off from $T_c$ by $2\pm1$\%
or more.  None of our $T>T_c$ runs showed any tunnelings between the
different ordered states except the run at $T/T_c=1.02$ on the $4 \times
16^3$ lattice, where we found one tunneling event between two deconfined
vacua (which forced us to use only the correlations of loop operators
here).  Within the precision of our measurement, the screening masses at
$T/T_c=1.02$ are independent of the lattice size: an indication that they
are not tunneling masses. For all our measurements of the $A_1^{++}$,
shown in Fig. \ref{fg.a1p}, the masses extracted from the Wilson loops
and those from the Wilson line agree at the 95\% confidence level.

The rough agreement between the measurement at $1.02T_c$ and that at $T_c$
in the deconfined phase, displayed in Figure \ref{fg.a1p}, should also be
noted.  We also draw attention to the feature that the screening masses,
when expressed in units of $T$, dip near $T_c$.  If the dip persists,
then it cannot be understood in the context of the perturbation theory
leading to eq.\ (\ref{pert}). An attempt to capture this effect in a
model has been made in \cite{rob}.  We plan a more detailed study in
both phases of the region near $T_c$.

In \cite{olaf} screening masses have been extracted from the free energy
change due to the addition of a static quark pair in the plasma, {\sl
i.e.\/}, the logarithm of the point-to-point Wilson-line correlation. Our
measurements of $m(T)/T$ for the scalar in the deconfined phase are
completely compatible with their results. This agreement is non-trivial
since we extract the screening mass from zero-momentum screening
correlators.

\begin{figure}[hbt]\begin{center}
   \scalebox{0.6}{\includegraphics{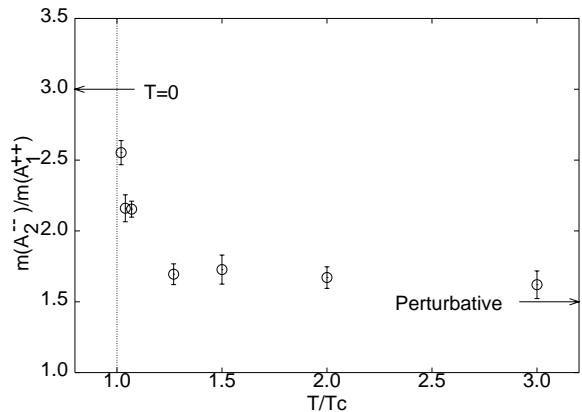}}
   \end{center}
   \caption{Ratio of the screening masses in the $A_2^{--}$ and $A_1^{++}$
      sectors as a function of temperature.}
\label{fg.ratio}\end{figure}

%Our results for the $A_2^{--}$ screening mass are also shown
%in Figure \ref{fg.a1p} \cite{owe}.  In Figure \ref{fg.ratio} we have
%displayed the same data in the form of the ratio of the two screening
%masses. Clearly, already at temperatures a little above $1.25T_c$, the
%ratio is close to the perturbative value of 3/2, approaching it from above.
%Earlier, by analysing relations between $A_1^{++}$, $B_1^{\pm+}$ and
%$B_2^{\pm+}$ screening masses, it was shown that no consistent meaning
%could be given to the mass of the magnetic gluons coming from the
%spatial links \cite{saumen}, thus protecting the argument which led to a
%prediction of 3/2 for the ratio. Even if we were to fly in the face of
%data and disregard this argument, the dominance of the electric gluon
%exchanges follows from our present observations through yet another
%simple argument.  Since the $B_1^{++}$ (which is much heavier than the
%scalar \cite{saumen}) must be due to the exchange of two magnetic gluons,
%and no combination of purely magnetic gluons project on the $A_2^{--}$,
%the ratio must be significantly less than 3/2 if magnetic gluons were
%freely propagating in the plasma. The very fact that the ratio is close
%to 3/2 shows that electric gluon exchanges dominate the correlator. Even
%where the ratio departs from this value, since it rises, it cannot be
%due to the presence of magnetic gluons.

Our results for the $A_2^{--}$ screening mass are also shown in Figure
\ref{fg.a1p} \cite{owe}.  In Figure \ref{fg.ratio} we have displayed the
same data in the form of the ratio of the two screening masses. Clearly,
already at temperatures a little above $1.25T_c$, the ratio is close to
the perturbative value of 3/2, approaching it from above.  It has been
shown earlier \cite{saumen} that no mass assignment for magnetic gluons
simultaneously satisfies the $B_1^{++}$ and $B_2^{++}$ screening masses,
both of which should arise due to the exchange of two such gluons. If
these colour singlet channels are treated as collective excitations,
then their large screening mass prevents them from contributing through
a pair exchange to the $A_1^{++}$ correlators.  Even if we choose to
disregard this argument, and insist on the existence of magnetic gluons,
their screening mass is very high (in this temperature range) simply
because of the large screening mass of the $B_1^{++}$. The simple fact
that the $A_1^{++}$ has the smallest screening mass protects the argument
of electric gluon exchange when the ratio $m(A_2^{--})/m(A_1^{++})$
is seen to be 3/2 as in Figure \ref{fg.ratio}.

We have not extended our measurement of the $A_2^{--}$ screening mass
below $T_c$ since previous studies have shown that in the confined
phase the degeneracies of the screening mass are those expected from the
$T=0$ symmetries of the transfer matrix \cite{bernd}. At $T=0$
the $A_2^{--}$ is just one component of a very heavy vector glueball
($T_1^{--}$ in the usual $T=0$ notation), and the measurement of
its screening mass would require large numbers of operators and a
significantly larger data set. It is interesting, though, that $T=0$
measurements, made over a broad range of lattice spacings, show that the
ratio of this vector mass to the scalar is roughly 3 \cite{peardon}.
The ratio $m(A_1^{++})/m(A_2^{--})$ for $T>T_c$ seems to interpolate
smoothly between the $T=0$ and the infinite temperature values, crossing
over rapidly from one regime to another in the temperature range between
$T_c$ and $1.25T_c$.

It is interesting to recall that the entropy density of the plasma
is relatively low for $T/T_c<1.1$, and begins to saturate only for
$T/T_c>1.25$ \cite{entropy}. While thermodynamics is more sensitive
to short distance modes and screening to long-distance modes in the
plasma, the two observations can nevertheless be due to a single cause,
if the partition function for $T/T_c<1.1$ were essentially saturated
by the colour singlet modes that we have seen. At sufficiently high
temperatures, the propagating electric gluons can contribute significantly
to the entropy \cite{entropy2}.

In summary, we have shown that the screening in the scalar ($A_1^{++}$)
sector is discontinuous across the pure QCD transition at $T_c$ and
falls abruptly by a factor of nearly 3 in going from the confined to
the deconfined phase. At larger temperatures, this screening mass rises
faster than linearly in $T$. Above $T_c$, the ratio of the $A_2^{--}$
to the scalar screening mass interpolates smoothly from a value close
to the $T=0$ ratio to the value 3/2 expected when these correlations
are saturated by electric gluon exchange. The agreement with the latter
value for $T>1.25T_c$ indicates the presence of weakly coupled, massive
quasi-particles with the quantum numbers of the electric gluon as the
lightest excitations of the QCD plasma.  Closer to $T_c$ the mass ratio
is more similar to its $T=0$ value. Taken together with the smallness
of the entropy density, this indicates that excitations are more nearly
glueball-like.

We would like to thank Rob Pisarski for correspondence on the temperature
variation of the $A_1^{++}$ screening mass. SD is supported by GSI under
project number BI-KAR.


\begin{thebibliography}{99}
\bibitem{irreps}
  $D_4$ has four one-dimensional irreps $A_{1,2}$ and $B_{1,2}$ and the
  two-dimensional irrep $E$. The signs in the superscripts label the
  irreps of $Z_2(T)$ and $Z_2(C)$ respectively. Each irrep of $D_4 \times
  Z_2(T)$ can be represented by some monomial in the coordinates $x$, $y$
  and $t$.  Even powers of $t$ correspond to $A_1^+$, and odd powers to
  $A_2^-$.
\bibitem{rotate}
  If the Wilson line settles into the wrong vacuum by chance, then this
  constraint can still be satisfied by simply implementing a global $Z_3$ rotation
  on the lines (SD thanks Sanatan Digal for this remark; see \cite{arnold} for
  more details). Of course, tunneling events cannot be handled so simply.
\end{thebibliography}
\end{document}